\pdfoutput=1
\documentclass[acmsmall]{acmart}

\AtBeginDocument{%
  \providecommand\BibTeX{{%
    \normalfont B\kern-0.5em{\scshape i\kern-0.25em b}\kern-0.8em\TeX}}}

\setcopyright{acmcopyright}
\copyrightyear{2020}
\acmYear{2020}
\acmDOI{10.1145/XXX.XXX}

\acmJournal{tochi}
\acmVolume{1}
\acmNumber{1}
\acmMonth{1}

\usepackage{balance}       
\usepackage{graphics}      
\usepackage{color}
\usepackage{booktabs}
\usepackage{multirow}
\usepackage{tabularx}

\usepackage{wrapfig,lipsum,booktabs}
\usepackage{xcolor}
\usepackage{subcaption}

\usepackage{colortbl}
\definecolor{accuracy_c}{RGB}{185,250,246}
\definecolor{automation_c}{RGB}{255,253,218}

\newcommand{\fullXai}{\textit{High Veracity Exp}}
\newcommand{\noai}{\textit{No AI}}
\newcommand{\randomXai}{\textit{Low Veracity Exp}}
\newcommand{\noXai}{\textit{No Exp}}

\newcommand{\revTask}{\textit{query review task}}

\newcommand{\randomShort}{\textit{Low Ver}}
\newcommand{\fullShort}{\textit{High Ver}}

\newcommand{\score}{\textit{component scores}}
\newcommand{\comb}{\textit{detected combination of components}}

\begin{document}

\title{Don't Explain without Verifying Veracity: An Evaluation of Explainable AI with Video Activity Recognition}

\author{Mahsan Nourani}
\email{mahsannourani@ufl.edu}
\affiliation{%
  \institution{University of Florida}
  \city{Gainesville}
  \state{Florida}
}
\author{Chiradeep Roy}
\email{cxr161630@utdallas.edu}
\affiliation{%
  \institution{University of Texas in Dallas}
  \city{Dallas}
  \state{Texas}
}
\author{Tahrima Rahman}
\email{tahrima.rahman@utdallas.edu}
\affiliation{%
  \institution{University of Texas in Dallas}
  \city{Dallas}
  \state{Texas}
}
\author{Eric D. Ragan}
\email{eragan@ufl.edu}
\affiliation{%
  \institution{University of Florida}
  \city{Gainesville}
  \state{Florida}
}
\author{Nicholas Ruozzi}
\email{nicholas.ruozzi@utdallas.edu}
\affiliation{%
  \institution{University of Texas in Dallas}
  \city{Dallas}
  \state{Texas}
}
\author{Vibhav Gogate}
\email{vibhav.gogate@utdallas.edu}
\affiliation{%
  \institution{University of Texas in Dallas}
  \city{Dallas}
  \state{Texas}
}

\renewcommand{\shortauthors}{Nourani et al.}

\begin{abstract}
Explainable machine learning and artificial intelligence models have been used to justify a model's decision-making process.
This added transparency aims to help improve user performance and understanding of the underlying model.
However, in practice, explainable systems face many open questions and challenges.
Specifically, designers might reduce the complexity of deep learning models in order to provide interpretability.
The explanations generated by these simplified models, however, might not accurately justify and be truthful to the model.
This can further add confusion to the users as they might not find the explanations meaningful with respect to the model predictions.
Understanding how these explanations affect user behavior is an ongoing challenge.
In this paper, we explore how explanation veracity affects user performance and agreement in intelligent systems.
Through a controlled user study with an explainable activity recognition system, we compare variations in explanation veracity for a video review and querying task.
The results suggest that low veracity explanations significantly decrease user performance and agreement compared to both accurate explanations and a system without explanations.
These findings demonstrate the importance of accurate and understandable explanations and caution that poor explanations can sometimes be worse than no explanations with respect to their effect on user performance and reliance on an AI system.

\end{abstract}



\maketitle


\section{Introduction}
\label{sec:Introduction}

Many machine learning models---especially those induced using deep learning approaches---are seen as black-boxes in that they do not allow users to understand ``why'' the system made a particular decision or produced a particular output \cite{krause2016interacting, marcus2018deep, holzinger2018machine}.
At a high level, in order to maximize  prediction accuracy, deep neural networks use a large number of hidden nodes where each hidden node represents a complex feature over the input attributes and other hidden nodes that it is connected to. 
A hidden node is activated when the feature it represents evaluates to ``true'' and thus in principle, it is possible to explain the decision made by a neural network by tracing the activated hidden nodes and constructing a (complex) feature over these nodes.
Unfortunately, since the feature associated with the hidden node is likely to be complex, it will not be human interpretable. 
As a result, neural networks are unable to generate meaningful, human interpretable explanations for their decisions \cite{marcus2018deep}. 
A lack of transparency can cause many problems for end users, as they would not know precisely how the outputs are generated, whether they are properly justified, and when they are wrong \cite{gui2017machine}.
These challenges have motivated interest in \textit{explainable artificial intelligence} (XAI), which seeks to bring more transparency to an intelligent system and make AI output and rationale more human understandable \cite{guidotti2019survey, tamagnini2017interpreting, holzinger2018machine}.


To date, a major focus of XAI research has been on introducing novel explanation types and then developing and evaluating models and algorithms for these new types \cite{lei-etal-2016-rationalizing,hendricks&al16,petsiuk2018rise}.
Unfortunately, little focus has been given to studying the effect of explanations, including their quality or lack thereof on end users. 
In fact, in a recent survey, Addadi and Berrada \shortcite{adadi2018peeking} argue that 95\% of the papers in the XAI community are focused on evaluating the accuracy of XAI algorithms rather than meaningful human use cases. 
As a result, advancements in XAI are overlooking fundamental knowledge of how explanations and human understanding of models affect user performance~\cite{hoffman2018metrics}, trust~\cite{cramer2008effects, dzindolet2003role}, and other behaviors when working with intelligent systems. 
This knowledge is crucial for designing effective XAI interfaces with meaningful, human understandable explanations, and to acquire such knowledge we need to conduct rigorous user studies to provide the empirical foundations for interpretability and trust in XAI systems~\cite{abdul2018trends}.


Effective evaluation of XAI systems is challenging because it must not only assess how the addition of explanations can improve user understanding and trust in the system but also whether improvements in understanding allows users to work more efficiently~\cite{mohseni2019survey}.
Human evaluation should also aim to understand which aspects or types of explanations aid human understanding\textemdash especially when several types of explanations are provided to the user.
Some machine learning models may be inherently harder to explain than others~\cite{goebel2018explainable,du2019techniques}.
Explanation designs are often approximate representations rather than fully-detailed or perfectly accurate representations.
It is important that explanations are truthful to the model in order to appropriately reflect the logic and rational behind its decision-making.
In this paper, we refer to this alignment between truthful and accurate reflection of machine as \textit{veracity}.
While important, explanation veracity is not necessarily sufficient for a understandable and meaningful explanations.
For end users, even explanations that accurately communicate model logic can sometimes be jarring if they cannot be interpreted as aligning with human logic or if they do not match expectations for how that logic would be portrayed~\cite{nourani2019effects, papenmeier2019model}.
Meaningfulness and truthfulness are sometimes distinct properties of an explanation (i.e., an explanation can be truthful but not meaningful, or vice versa).
Although, in other cases, this truthfulness can affect how meaningful users find the explanations, as well as users' understanding of the model.
In this paper, we explore such phenomena through a study to better understand the impact of explanation veracity for end users.

We study how the veracity of explanations can affect user task performance and system understanding through a controlled experiment conducted with an explainable activity recognition system for video.
To circumvent the poor interpretability of neural networks, our system is built on a probabilistic model that takes the output of the neural network as input and models relationships between entities recognized by the neural network. 
The main virtue of this approach is that we can generate high quality explanations by reasoning over the probabilistic model~\cite{roy2019explainable}, and this method provides an explainable model as the basis for human-subjects evaluation. 
By conducting an in-depth user evaluation with this XAI model, we are able to maintain experimental control of the presence and nature of available explanation information for our evaluation.
At the same time, evaluating with a practical XAI implementation grounds the research in a real system with capabilities similar to those desired for real-world applications of activity recognition, e.g., security monitoring, medical analysis, or disaster response.

The results of this study contribute empirical evidence of how high-veracity explanations can benefit human-machine operations while also validating concerns about the risks of inaccurate or non-meaningful explanations.

\section{Related Work}
\label{sec:RelatedWork}

In this section, we discuss the current body of work in the explainable AI literature with two motivations, one focusing on generating, refining, and combining AI/ML algorithms and the other, focusing on evaluating and studying the existing systems and models from a human viewpoint.

\subsection{Explainable Machine Learning and Artificial Intelligence}

Over the last two decades, machine learning and artificial intelligence (ML/AI) have fundamentally changed user experience by making computer systems smarter and more intelligent. 
In fact, a large number of existing ML/AI systems have achieved relative autonomy in that they can decide and act on their own with minimal human intervention. 
However, a major limitation of these existing systems is that they are black-boxes and cannot explain why they made a particular decision. 
As a result, users who interact with these autonomous systems on a daily basis are unable to understand and trust them, especially when they make a counter-intuitive decision. 
Therefore, recently there has been growing interest in building explainable artificial intelligence (XAI)\textemdash specifically explainable machine learning\textemdash systems.
Notable examples include \textit{explainable} recommendation systems (e.g.,~\cite{wang2018tem, cheng2019mmalfm, zhang2014explicit}), classification systems (e.g.,~\cite{ribeiro2016should, kim2019unsupervised, alonso2018explainable, wang2019deepvid}), and activity recognition systems (e.g.,~\cite{zhou2018temporal, roy2019explainable, meng2018and,atzmueller2018explicative}).
In this paper, we focused on an explainable activity recognition system for videos based on our previous work~\cite{roy2019explainable}.

Various researchers in the ML/AI communities have explored explainability and interpretability techniques in ML models and systems.
For instance, Du et al.~\shortcite{du2019techniques} present a survey on different interpretability techniques, including post-hoc explanations (where explanations are extracted from the model from local or global perspectives, e.g., \cite{guidotti2018survey,laugel2019dangers,korber2018have}), intrinsic explanations (where the model is self-explanatory and is built to be interpretable globally or locally, e.g.,~\cite{vandewiele2016genesim}), and model specific/agnostic explanations (e.g., \cite{ribeiro2018anchors}).
Other researchers have explored explanation by example, where a model provides examples of relevant instances from the training set for a given input instead of attempting to explicitly explaining how the model's logic.
For instance, Cai, Jongejan, and Holbrook~\shortcite{cai2019effects} defined and explored two types of example-based explanations in the visual domain and investigated their effectiveness with humans: (1) normative explanations establish a norm/trend for the target class by showing training examples, which would help the users understand classifications, and (2) comparative explanations show the most similar examples (which can be of different classifications) from the training set to the input.

Researchers also describe types of focus for what is being explained by different explanations.
For example, Keane and Kenny~\shortcite{keane2019case} argue that \textit{transparency} tries to reflect \textit{how} an AI system works, while \textit{post-hoc} interpretability focuses on the \textit{why}s in the AI system, providing justification for its outputs.
In another work, Hohman et al.~\shortcite{hohman2018visual} provide an interrogative survey on a large number of works in Deep Neural Network (DNN) visualization papers and organize the literature into six categorise based on how the visualization could reveal different aspects of a DNN.
These categorise are: (1) \textit{Why} visualize deep learning models? (e.g.,~\cite{montavon2018methods,lipton2018mythos}), (2) \textit{What} data, features, and relationships can be visualized? (e.g.,\cite{harley2015interactive}), (3) \textit{When} is visualization used in deep learning? (e.g.,~\cite{pezzotti2017deepeyes,kahng2017cti}), (4) \textit{Who} would use and benefit from visualization of deep learning?, (5) \textit{How} to visualize data, features, and relationships? (e.g.,~\cite{kahng2017cti}), and (6) \textit{Where} has the visualization of deep learning been used? (e.g.,~\cite{robinson2017deep,zahavy2016graying}.
While these categories are presented for visualization of deep learning approaches, similar questions are relevant for all interpretable and explainable AI systems.

In our work presented in this paper,  our explainable system follows a similar approach and targets a critical explainable system (explainable video activity recognition), though it differs from much of the work from machine learning community in that we aim to evaluate and understand user behaviors with explainable systems rather than focus on improving the model itself.
Next, we will discuss work on strategies for designing \textit{how} to explain to human users and how explanations can affect humans.

\subsection{Design and Human Factors in Explainable AI}

A variety of different research communities have studied the design of user-centered XAI systems.
In the visualization community, researchers have focused on building and designing tools for XAI systems in an attempt to improve understanding~\cite{spinner2019explainer}, analytical process~\cite{pena2019detecting, zhang2018manifold}, debugging~\cite{strobelt2018s, zhang2018manifold}, and fairness~\cite{ahn2019fairsight}.
Work in the HCI community has resulted in user-centered design guidelines and frameworks for XAI systems
~\cite{eiband2018bringing, wang2019designing, wolf2019explainability}, while others have compared and evaluated design variations for explanations to understand how they can improve user experience in an XAI system~\cite{keil2006explanation, lombrozo2007simplicity}.
Some HCI researchers focused on building empirical knowledge around explainable intelligent systems by studying user behaviors with such systems.
Evaluation of user mental models of intelligent systems is one the topics explored in the HCI community.
In the context of intelligent systems, mental model refers to the user's built and formed concept of how a system works in their mind from a period of working with the system.
In 1993, Staggers and Norcio~\shortcite{staggers1993mental} argued that most of the research attempts infer the presence of mental models by comparing users' performances and observing differences in problem-solving between novice and expert users within a certain domain; however, it is hard in general to measure mental models.
Some researchers focused on studying how transparency can improve user mental model of an intelligent system.
Eiband et al.~\shortcite{eiband2018bringing} proposed guidelines on how to improve the transparency in intelligent user interfaces by studying expert, user, and target mental models iteratively to understand what to explain to the users.
More recently, Hoffman et al.~\shortcite{hoffman2018metrics} has proposed several different metrics such as trust, user-machine task performance, and mental model for user evaluation in XAI systems.  
In his paper, he lists multiple techniques to elicit mental models in XAI systems, e.g., \textit{think-aloud problem solving}, \textit{card sorting}, and \textit{prediction task} (where users are presented with a test case and asked to predict the results of the system given the input).
For instance, Kulesza et al. \shortcite{kulesza2013too}  studied how explanation soundness and completeness affect user mental model through the \textit{think-aloud approach}.
Here, we adapt a prediction task as used in several existing XAI user evaluations, e.g., \cite{nourani2019effects, anderson2019explaining, madumal2019explainable}, to assess user mental model.
In other work, Nourani et al.~\shortcite{Nourani2020Primacy} studied the order of observed system output and explanations on user mental model and found that order of encountering accurate or erroneous system predictions has a more significant effect on user mental model and performance than explanations.

Previous research in HCI has also been conducted to understand the relationship between user mental model and performance \cite{xie2017influential}.
Some work has shown that a proper mental model of a system can have positive effects on user performance \cite{dimitroff1992mental,ziefle2004mental}, while other work shows a contradicting effect \cite{norman1983design,borgman1986user}. 
Overall, the existing body of work in HCI and psychology does provide evidence that user performance and mental model are correlated \cite{xie2017influential}.
How they are correlated, however, could depend on  task, context, study population, or other external factors~\cite{10.1007/978-3-319-58536-9_45}.

Helping users build a proper mental model in intelligent systems is critical for various reasons.
One goal, which also aligns with a goal of adding transparency to intelligent systems, is to help the users' trust-building with the system \cite{garcia2018explainable}.
For instance, Yin et al. \shortcite{yin2019understanding} found that users' observed system accuracy in intelligent systems can significantly affect user trust.
While adding transparency and explanations can improve user trust in intelligent systems, it can introduce further issues with user behaviors. 
Bussone et al.~\shortcite{bussone2015role} studied the relationships between explanations and trust and reliance through a study facilitated by a clinical decision support system.
Their results demonstrate that more detailed explanations can improve user trust, though may also lead to over-reliance issues.
However, given less detailed explanations had a counter effect and caused self-reliance.

Either over-reliance or under-reliance can affect the quality of user experience and more importantly, could cause fatal errors in  decision-making processes.
One common example of over-reliance is \textit{automation bias} \cite{cummings2004automation}, a psychological bias where users sometimes tend to default to trusting and relying on the outputs of an intelligent system when they become comfortable with the system's performance or think it ``seems smart''.
Automation bias can cause problems such as subconsciously ignoring or missing system errors (omission), and ignoring the contradictory factors in the decision-making process and following system suggestions (commission) \cite{parasuraman1997humans,cummings2004automation}.
On the other hand, self-reliance and mistrust in an automated system can cause disuse, inadequate user performance, or disuse \cite{parasuraman1997humans}.
Both over-reliance (\textit{misuse}) and self-reliance (\textit{disuse}) are of particularly relevant as many intelligent systems target novice users, who might bring incorrect assumptions about the capabilities of intelligent systems \cite{janssen2019history}.
Numerous studies have investigated automation bias in XAI systems.
For example, Schaffer et al. \shortcite{schaffer2019can}
designed an experiment to test how presence of explanations, level of automation, and level of system error influences user's acceptance of advice from the intelligent system.
Their results show explanations helped people with less task familiarity more, while showing explanations to users with higher task familiarity led to automation bias.
Furthermore, Lim and Dey \shortcite{lim2011investigating} found that more descriptive, detailed explanations can lead users to disagree with the system.
This was an opposite effect from automation bias and might be helpful in avoiding such scenarios.
In our experiment, we focus on learning more about how explanation quality and understandability can affect user agreement with the system and whether any effects interact with user reliance and trust.

\subsection{Explanation Meaningfulness and Veracity}

Explanations come in different formats, such as textual~\cite{bussone2015role}, confidence scores~\cite{bussone2015role} and prediction accuracies~\cite{schaffer2019can}, and saliency maps~\cite{dabkowski2017real,ribeiro2018anchors}.
They can also cover different scopes of model operation and logic.
\textit{Global} explanations try to provide an overview of how a model generates its outputs~\cite{hohman2019s,adadi2018peeking}.
Many explainable systems designed for data experts have focused on visualizing models as global explanations~\cite{mohseni2019survey}.
For instance, Hohman et al.~\shortcite{hohman2019s} built an interactive visual system to summarize and visualize deep-learning models and show how much each layer and what features were used to make predictions.
Global explanations are beneficial as they might reveal biases, help diagnose model problems, and allow potential for changing hyper parameters \cite{ibrahim2019global}.
A downside of global explanations is they are harder to achieve in practice, especially for complex or deep learning models~\cite{adadi2018peeking}.
In contrast to global explanation, \textit{local} explanation aims to provide justification for particular outputs given specific instances of input.
For instance, Bussone et al.~\shortcite{bussone2015role} show users why a certain diagnosis was suggested based on the patient's symptoms and medical history.
An individual local explanation generally does not give an overall view of the how the model works, but users can gain an understanding as they continue working with the system over time and view multiple instances. 
Depending on the task, instance-level explanation can be critical in aiding user's understanding of system output.
Our study uses local explanations as we focus on explaining relevant features of instances responsible for an output, and the goal was to study system and scenario that did not require any particular data expertise. 

One specific problem with local explanations is how they can affect user's perception of the system  quickly.
Hoffman~\shortcite{hoffman2013trust} noted that it is easy to lose trust in automation, but harder to reestablish trust once it is lost.
For that matter, poor or incorrect instance-level explanations can cause an immediate loss of trust.
Hence, the quality of these explanations is of high importance.

Researchers address aspects of ``quality'' of explanation from different perspectives.
Recently, Papenmeier et al.~\shortcite{papenmeier2019model} investigated how explanation presence and fidelity (i.e., ``how truthfully the explanation represents the underlying model'') and system accuracy influence user trust.
Utilizing two levels of high and low fidelity explanations, they found that system accuracy plays an important role in building user trust, while low fidelity explanations can potentially harm trust.
Other work has explored ``nonsensical explanations'', which by their definition, mean explanations that users cannot make sense of.
One of the earliest uses of this term was by Langer et al.~\shortcite{langer1978mindlessness}, when three behavioural field experiments led to the finding that people adhere to an explanation when it is more informative rather than senseless (nonsensical).
More recently, Feng et al.~\shortcite{feng2018pathologies} introduced a method of input reduction in their neural model that reduced explanations by removing unimportant words while maintaining its accuracy.
However, their human evaluation showed that these shortened explanations and summaries confuse humans as they found them nonsensical, resulting in a drop in their task accuracy.

Other research also considers how \textit{meaningful} explanations are for users.
By some interpretations, explanations can be considered meaningful if they can be understood in alignment with human understanding and logic~\cite{nourani2019effects}.
Hind et al.~\shortcite{hind2019ted} argue that it is not possible to provide one single instance of explanation that is meaningful to everyone, as some factors regarding the users, such as their domain knowledge and level of sophistication, play an important role in designing explanations.
They introduced a framework, so-called TED (Teaching Explanations for Decisions) to provide meaningful explanations that matches user mental models while maintaining prediction accuracy for the algorithm.
Codella et al~\shortcite{codella2019teaching} explored and evaluated the TED framework further and demonstrated that machine learning approaches can incorporate meaningful explanations reliably, and in some cases, these explanations can be used to improve model accuracy.

In our prior work \shortcite{nourani2019effects}, we conducted an experiment to investigate the role of human-meaningfulness in explanations for influencing users' perception of model accuracy.
Using a wizard-of-Oz study with an image classification task, the study compared meaningful explanations with nonsensical or meaningless explanations intentionally created to misalign with natural human interpretation.
The experiment compared meaningful and meaningless explanations alongside a control with no explanations, and it also controlled two levels of system accuracy through between-subjects user study.
After viewing repeated instances of the (fictitious) system's classification for sample input images, participants performed a prediction task (see \cite{hoffman2018metrics}) for a new set of images.
The guessed accuracy from the prediction task was used as an implicit measure for perceived system accuracy, and participants also provided an explicit numerical estimation.
For both measures of perception of accuracy, the results showed participants who are exposed to the nonsensical explanations significantly underestimated accuracy compared to both users with no explanations and meaningful explanations.
These results---taken along with existing empirical evidence that user perception of accuracy can directly affect user trust~\cite{yin2019understanding}---suggest explanation meaningfulness can significantly affect user trust.


In the current paper, we expand on the study of meaningful explanations with an experiment using a real system rather than a simulated scenario, and we also account for additional user outcomes such as user-task performance (as assessed in many XAI studies relevant work~\cite{lim2009and,lim2011investigating,rovira2007effects}).
Highly relevant to meaningfulness is the concept of \textit{veracity} (also called soundness), which we use to refer to the extent that explanations are truthful to the reality of the system or model they explain~\cite{kulesza2013too,staron2016data}.
Veracity differs from explanation meaningfulness in that veracity is based on correctness in comparison to the underlying model, whereas meaningfulness is based on comparison to human logic or thinking.

Studying veracity in transportation systems as a component of big data, Staron and Scandariato~\shortcite{staron2016data} defined veracity as how data is truthful compared to reality, or the ability of the data ``to be free from lies''.
They believe data veracity is an important factor for relying on data accuracy and truthfulness.
Thus, they define a list of characteristics where one could verify the veracity in big data.
Other studies of veracity have investigated similar perspectives for information and data accuracy.
For instance, Levine~\shortcite{levine2019overview} defines \textit{truth accuracy} as the ability of humans to detect honest information within a collection of information, and \textit{lie accuracy} as the human ability to detect lies correctly in this collection.
The author describes \textit{truth-bias}, a phenomenon in which a human tends to judge information more honest when the truth accuracy is higher than the lie accuracy, and calls it a \textit{veracity effect}.
This may be similar to automation bias in intelligent systems.




\section{Explainable Video Activity Recognition System}
\label{sec:xaisystem}

Our experiments were conducted using a custom-developed explainable activity recognition system for video.
Activity recognition is an ideal test bed for XAI research due to its many potential real-world applications (e.g., fire detection \cite{lai2012robust}, airport security \cite{tripathi2018suspicious}, smart hospitals \cite{sanchez2008activity, yeung2019computer}, and elderly care \cite{khan2011abnormal}) and because most activity recognition systems involve a substantial human-computer interaction component.
Our goal is to study system understanding and system effectiveness among a non-specialist population, i.e., users without any particular domain expertise or AI knowledge.
That is why we chose to design a system to \textit{identify cooking activities in a kitchen setting}.

Our system outputs human understandable explanations using a two-layer architecture with a tractable, interpretable model on top of a deep, uninterpretable layer. 
We note that we presented a preliminary version of the system in an earlier workshop paper \cite{roy2019explainable}, but the current paper is based on an updated system and a completely new evaluation. 
In this section, we describe the dataset, model, and interface for the explainable activity recognition system used in the evaluation.
For convenience, in the remainder of this paper, we will refer to this system as the XAI system.

\subsection{Dataset}

The model for the activity detection system was trained using a publicly-available video dataset, the Textually Annotated Cooking Scenes (TACoS) dataset~\cite{regneri2013grounding}, which consists of videos of a many different cooking-related activities. 
For example, a typical video will have a person take out a vegetable from the refrigerator, wash it, cut it, and then cook it.
The cooking context has the advantage of being easily understandable, even without particular domain expertise.
The dataset includes hand-annotated labels of actions, objects, and locations for each frame of video. We isolate 28 such labels and use videos with only these labels for our experiments.
Most videos are around 2 minutes in length (although videos as long as 15 minutes are also present).
For training, we used 60313 frames divided over 17 videos. 
For testing, we used a different set of 9355 frames over the same 17 videos.


\subsection{Model}
This section describes the model that was used for our experiments. We use a two-layered architecture whose high-level overview can be found in Figure \ref{fig:fig1}.  More complete system details can be found in the supplementary material.

\subsubsection{Video Classification Layer}
This layer takes a frame (image) as input and outputs a 0/1 label for each activity (1 indicates that the activity happened while a 0 indicates that it did not) in the frame.
It is implemented using a 22-layer deep CNN architecture based on the BAIR/BLVC GoogleNet model \cite{szegedy2015going}.
This model is pre-trained \cite{szegedy2014pretrain} on the dataset used for the ILSVRC'15 object detection challenge \cite{ILSVRC15}. 
This dataset contains over 500,000 images of different objects from around 200 categories.
We added a soft-max layer on top that contains 28 nodes for our labels and then re-trained the network for our labeled data, i.e., the model was trained using the TaCOS video frames as inputs and the annotated ground labels as the desired outputs.
The accuracy of the model with respect to the test set was measured using information retrieval metrics such as the Jaccard Index (higher is better) and the Hamming Loss (smaller is better). The video classification does not model temporal relationships between frames. Nevertheless, it yields a competitive Jaccard Index of 0.8608 and hamming loss of 0.1392.

\begin{figure}[t]
  \centering
  \captionsetup{justification=centering}
  \includegraphics[width=0.9\linewidth]{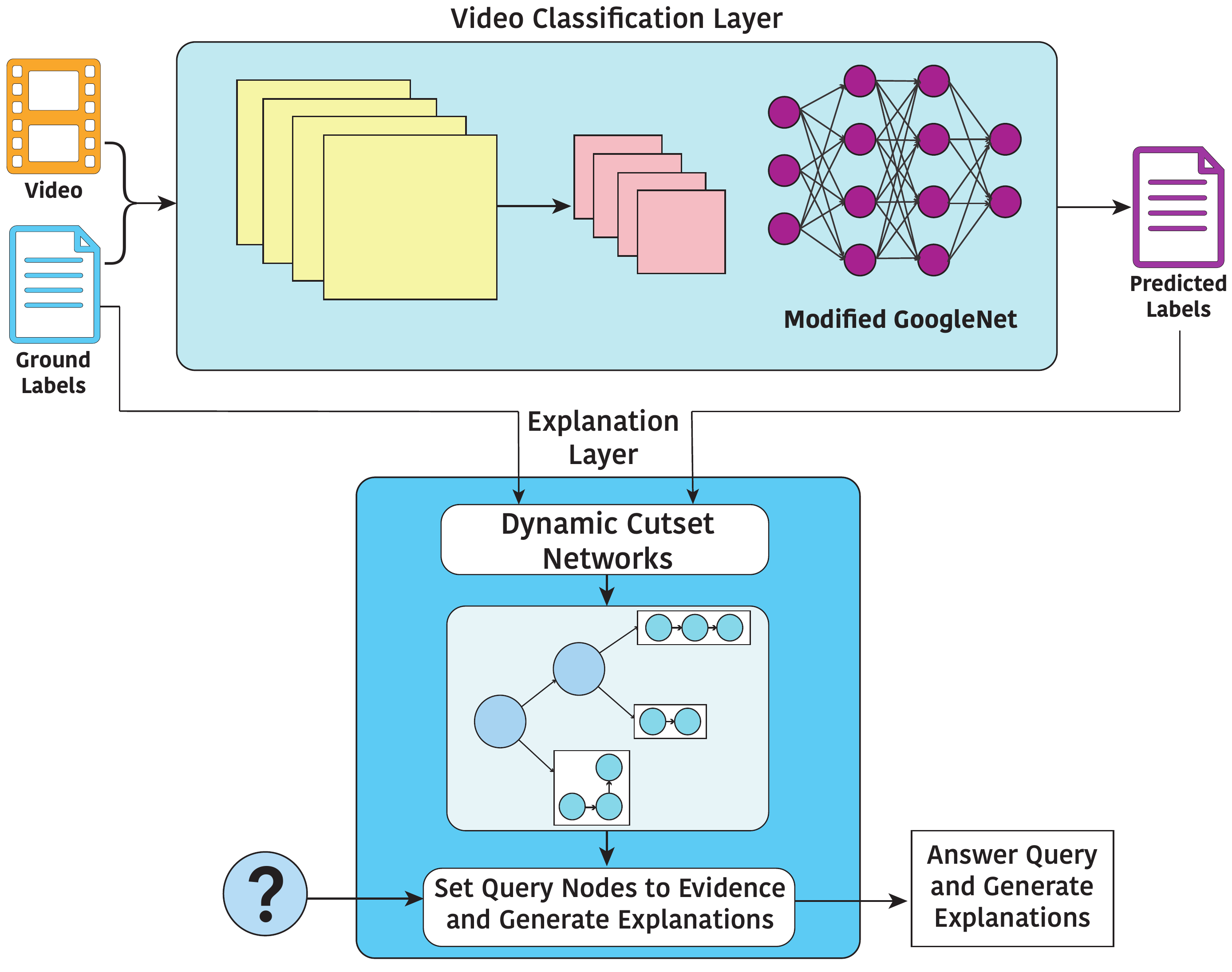}
  \caption{High-level architecture and data processing pipeline.}~\label{fig:fig1}
\end{figure}

\subsubsection{Explanation Layer}

The explanation layer consists of two parts:

  \begin{enumerate}
  
      \item \textbf{Training:} The predicted labels from the Video Classification Layer are used to train a temporal Dynamic Conditional Cutset Network (DCCN).
      Similar to Dynamic Bayesian Networks (DBNs) \cite{murphy2002dynamic}, DCCNs are temporal probabilistic models whose emission and transition distributions are represented using Conditional Cutset Networks \cite{rahman2014cutset}, instead of the Bayesian Networks used in DBNs.
      Since this layer models the temporal relationship between the predicted labels and the ground truth labels, it enables us to answer complex temporal inference queries and explain the answers via probabilistic reasoning. It also achieves a higher Jacard index (0.8687) and a smaller Hamming loss (0.1200) than the deep net alone. 
      \item \textbf{Query Resolution:} 
      The system allows users to enter a query about an activity composed of an \textit{action}, \textit{object}, and \textit{location}.
      For example, a query of ``Did the person cut the orange on the plate?'' can be broken down into its corresponding activity tuple of (\textit{cut}, \textit{orange}, \textit{plate}).
      The system then uses the explanation layer to search for frame segments that are the most likely to contain the activity that we are searching for. 
      It also generates the top-k likely activities for each frame segment and ranks them in descending order of likelihood score (i.e., the most likely explanation will be ranked highest).
      The explanation information from this layer can then be presented in the system's visual interface.
  \end{enumerate}

\subsection{Interface}
\label{subsec:interface}


\begin{figure*}[t]
    \centering
    \includegraphics[width=\linewidth]{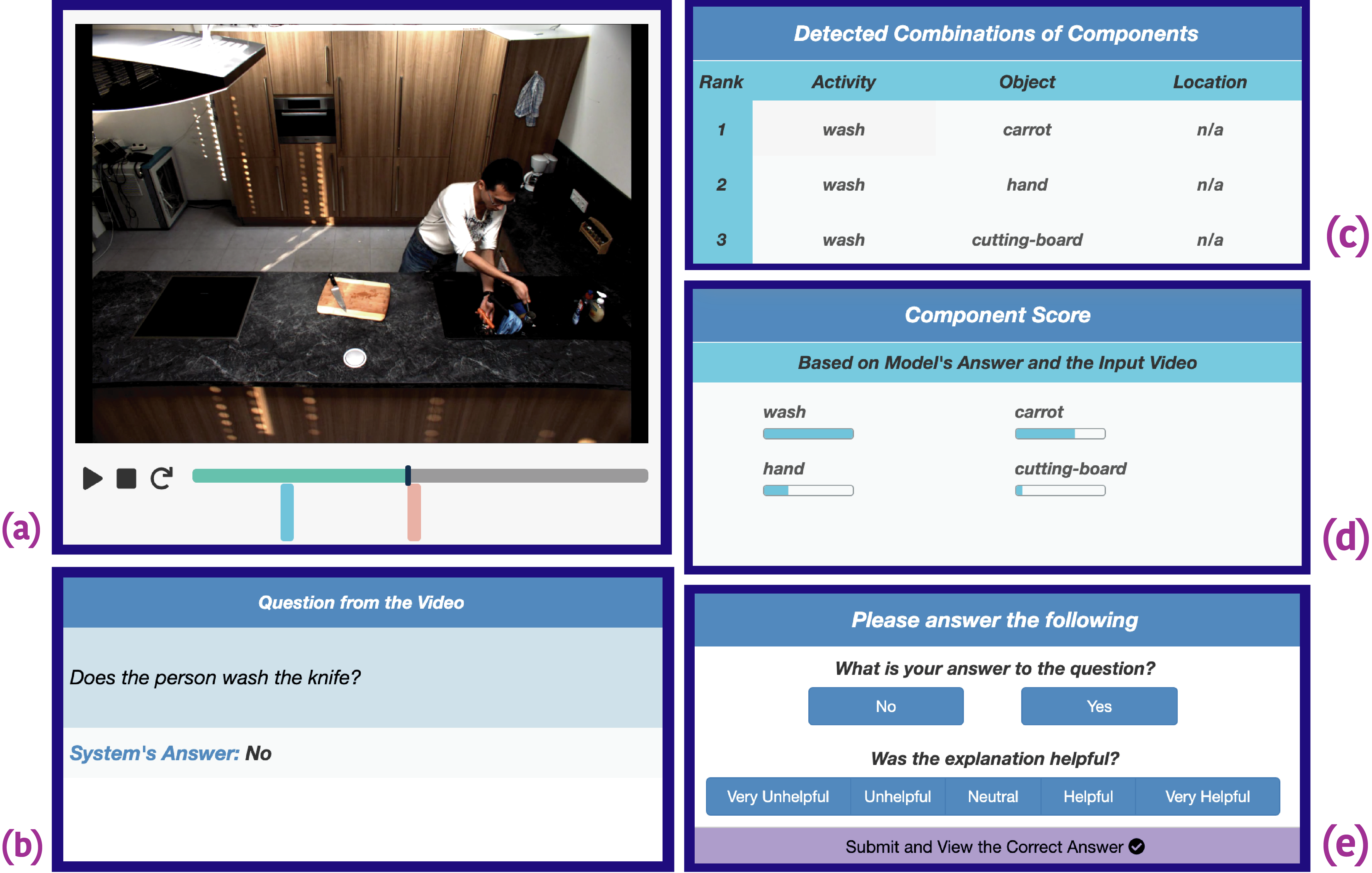}
    \caption{The elements designed and used for the experiment. Figure \ref{fig:interface}a shows the video panel, including media player and explanation segments. Figure \ref{fig:interface}b shows the query panel, including the trial-specific query alongside the system's answer and the correct answer (not shown until users submits their response). Figure \ref{fig:interface}c and \ref{fig:interface}d show the component scores and combinations based on the selected segment on the video panel. Figure \ref{fig:interface}e shows the question panel where the user answers questions for the trial.
    }
    \label{fig:interface}
\end{figure*}

The interface was designed to communicate the most relevant explanatory elements from the XAI activity recognition model for each given query and video input.
The system was implemented with an interactive web-based interface.
Figure~\ref{fig:interface} shows the components of the interface from the study (though available elements depended on the experimental conditions, as described in the following sections).

To allow regular video viewing, the interface included a custom a video player with typical functionality (e.g., play/pause, stop, replay) and an interactive progress bar for jumping to any particular play time (see Figure~\ref{fig:interface}a).
To help explain to users the relevant segments of the video that contributed to the XAI answer for each query, the video player showed key video segments directly under the progress bar.
The width of each segment bar showed the length of the relevant segment at the appropriate time in the video.
Figure~\ref{fig:interface}a shows two segment bars for the query shown.
Clicking on a segment bar would jump the video and progress bar to the start of that period of the video.  
Different queries could have different numbers of segments, where the currently-selected segment bar would be shown in orange to distinguish from the default blue color.
Upon submitting a new query, the first segment of the video was selected by default.

For each selected segment, the system also displays more detailed explanation information from the model. 
Explanations included \score{}, which are probability scores ranging from 0 to 1, which specify the contribution of every individual label (i.e., action, object, or location), in a segment for the most likely explanation from the explanation layer.
Scores are shown graphically with simple bar representations (see Figure~\ref{fig:interface}c).
Additionally, the top three most likely combinations of components together are shown in a list of \comb{} ordered from the most likely explanation at the top to the least likely at the bottom (see Figure~\ref{fig:interface}d).


While the full version of the system can process any user-entered query about actions, objects, and locations, the available functionality of the application was limited for purposes of the study.
To ensure experimental control and consistency for what queries and explanations participants encountered in the user study, queries were pre-determined with a single yes-or-no query given per trial (see Figure~\ref{fig:interface}b).


\section{Experiment}
\label{sec:Experiment}
We conducted a controlled experiment to study how the inclusion and veracity of explanations in explainable systems affect user performance, mental model of the system, and user perception of system accuracy. 
The study was run using the system described in Section \ref{sec:xaisystem}.
In this section, we describe the experimental design and evaluation methodology.

\subsection{Research Goals and Hypotheses}
\label{subsec:goals}

The primary motivation for this study was to understand how explanation veracity affects user performance and agreement with intelligent systems.
Since one of the main goals of adding explanations to an intelligent system to improve understanding and, ultimately, to enhance user performance~\cite{hoffman2018metrics} (i.e., allowing users to complete a task faster and with less error), our evaluation prioritizes assessment of human task performance with the assistance of the system.

We summarize our goals with three research questions:
\begin{itemize}
\item \textbf{RQ1}: Does the explainable AI model with full explanations improve user performance?
\item \textbf{RQ2}: How do presence and veracity of explanations affect user performance and user agreement with model output?
\item \textbf{RQ3}: How do presence and veracity of explanations affect perceived accuracy and mental model of intelligent systems?
\end{itemize}

While RQ1 is dependent on the specifics of the XAI implementation, we argue for the importance of evaluating specific models and explanation designs before using those systems as the basis for investigating broader research questions, such as our questions about veracity.
Furthermore, establishing empirical knowledge about the effects of explanations over a variety of systems and domains is also necessary to advance generalizable understanding of XAI effectiveness over broad contexts.
Assessment of the effectiveness of our particular XAI system is also important when considering the additional research questions RQ2 and RQ3.
If the explainable model used in the study is not effective and does not improve user performance, we cannot be confident that the design was sufficient or appropriate to provide a basis for comparison with system alternatives.
We therefore consider RQ1 to provide a baseline of XAI effectiveness with the intended design, and we hypothesized that the XAI system would significantly improve user performance compared to when no explanations or no AI model answers are provided.

RQ2 and RQ3 focus on effects of the level of veracity of provided explanations.
For RQ2, we hypothesized that lower veracity explanations would cause performance penalties compared to high veracity explanations.
We also hypothesized that users would exhibit greater agreement of decisions with system answers when provided high veracity explanations than with low veracity explanations.
We also questioned whether agreement without explanations would differ from cases with explanations since there is the possibility that the presence of \textit{any} explanations could potentially persuade users to think that the system is functioning intelligently and correctly (that is, more details might appear impressive and convincing, thus artificially improving perception of machine ability).

Finally, for RQ3, we hypothesized that users would have a higher perception of model accuracy and be able to more reliably predict system correctness when given higher veracity explanations compared to poor or no explanations. 
We also hypothesized that users with poor explanations might underestimate system accuracy significantly more than users with high veracity explanations.
This hypothesis is informed by results observed in our previous user study~\shortcite{nourani2019effects} using a simple classification scenario and a wizard-of-oz approach, but our study investigates the extension to an actual XAI model with explanation veracity (rather than explanation meaningfulness) in a full system implementation.

\subsection{Experimental Design}

To address our research goals, we designed an experiment around a user task requiring participants to use the activity recognition system to review given queries about videos.

\subsubsection{User Task}
\label{subsec:userTask}

The core of the experiment involved two consecutive tasks: a \revTask{} and a user \textit{prediction task}.
The \revTask{} was designed to assess whether the system affected how well participants could accurately determine the correct answer to queries with the aid of the system. 
Each trial involved review of a unique yes-no query about whether certain activities or content could be found in a given video of cooking activity from the TaCOS corpus.

In addition to providing user performance data for query assessment, the \revTask{} also gave participants a chance to develop a mental model of the system from the period of use with multiple queries. 
After this period, users moved on to a \textit{prediction task} to provide a measure of participant's mental model.
For this task, participants were given a new set of queries and new videos, but the system did not produce output (neither query answers nor explanations).
Participants were asked to predict what the model would answer for each query.
The prediction task (as suggested by others~\cite{hoffman2018metrics}) is a method for assessing a user's mental model of the system based on the ability to predict cases where the model provides correct answers and where it fails.
Additionally, the predicted ratio of correct and incorrect responses provides an implicit approximation of the user's perceived model accuracy (as similarly done in our previous study~\shortcite{nourani2019effects}).

\subsubsection{Conditions}

The study followed a between-subject design to control  XAI system output as a single independent variable with four conditions: 

\begin{itemize}
	
	\item \noai{}: The system did not provide the AI answers for the queries, and no explanatory information was given. 
	The system did allow full access to the video player to inspect videos for queries, but without relevant segment highlights or component information. 
	This was the only condition that did not include the AI output, and it was included as a baseline reference for human performance of the \revTask{}.
	
	\item \noXai{}: The system did show the AI answer to the query, but no explanatory information was given (i.e., no video segment highlights or component information).
	This condition served as a reference for human performance with the help of the AI answers to queries but without explanations.
	
	\item \fullXai{}: All intended functionality was provided as shown in Figure~\ref{fig:interface}. The system provided the AI answer from the model, along with the appropriate corresponding explanatory information (segment highlighting and component information) for the model's computation of that answer.
	
	\item \randomXai{}: All intended functionality was provided as shown in Figure~\ref{fig:interface}, and the system explanations for our study of explanation veracity.
	However, the explanatory information shown in this condition was not correct for the query.
	Model explanations (segment highlighting and component information) was generated based on other queries to simulate inaccurate explanations.

\end{itemize}

The experimental conditions only affected the version of the system available during the \revTask{} but the \textit{prediction task} was not affected since this was done to record data about user understanding and perception of the model; participants in all AI conditions completed the prediction task without AI query answers and without explanations.
We also clarify that all conditions included use of the video player, but the highlighted segments were only included under the play bar for conditions with explanations (\fullXai{} and \randomXai{}).

We also note that both the \fullXai{} and \randomXai{} conditions included the same AI query answers from the model (i.e., the AI accuracy was identical), but the difference was the veracity of the explanations.
While the \fullXai{} condition showed the actual intended XAI explanations as generated by the model, the purpose of the \randomXai{} condition was to show inaccurate explanation information.
To achieve this, incorrect explanations were simulated in the \randomXai{} condition by using explanations from the model for different queries.
That is, we mixed-and-matched explanations for and among queries that were used for the study.
We also used a priori experimenter review to make sure the assigned explanation was not obviously helpful or serendipitously highlighting relevant video content.

\subsubsection{Measures}

Outcomes for the experiment included participant performance from the \revTask{} in terms of completion time and error while determining the correct query answers with the aid of the system.
Responses were recorded per trial and measures were calculated as the average of all trials per participant. 

To evaluate perceived usefulness and willingness to rely on model outputs, we considered whether participants' answers for the queries matched the system's given answer.
The expectation is that participants demonstrating more trust and reliance in the system would be more likely to use the system's answer as their own answer.
Additionally, for the two study conditions that included explanations, participants rated explanation usefulness on a 5-point scale from ``very unhelpful'' to ``very helpful'' (see Figure~\ref{fig:interface}e).

To help assess perception and mental model of the system's AI, percentage of correct predictions from all trials of the \textit{prediction task} served as an indicator of participant understanding of the system.
In addition, the percentage of guessed correct/incorrect responses was used as an implicit measure of how accurate the participant perceived the system to be.
Participants also provided a direct numerical estimation of system accuracy in a post-study questionnaire.


\subsubsection{Query and Data Configuration}

The \revTask{} and \textit{prediction task} each had 20 queries (one query per trial).
In each task, participants reviewed four unique videos with five queries for each video.
In both the \revTask{} and \textit{prediction task}, participants completed the task for all queries one at a time for each video, though the order of queries for each video was randomly determined per participant.
In addition, the video order was also randomized.

To control the question composition to avoid any confounds, we used the same set of queries for all the conditions.
Since our research goals included evaluating whether participants agreed with system results and could accurately assess model accuracy, it was necessary to ensure that all participants observed examples of the cases where the system produced both correct and incorrect answers to queries in the \revTask{}.
We therefore composed the set of queries for the study such that participants would observe incorrect outputs more often than would normally be expected (in other words, we artificially reduced the accuracy by intentionally controlling the number of queries with right and wrong model outputs).
Of the 20 queries, the system gave correct query answers for 16 (i.e., 80\% accuracy).
This use of a constant, simulated accuracy provided consistency and ensured all participants had a sufficient number of observations of the system giving wrong and right answers.

To assist with quality assurance for the online study, we also included two additional trials (with the existing videos) as attention checks during the \revTask{}.
These trials affected neither the question composition and controlled system accuracy nor the final results.
These attention checks were simple questions that did not require inspection of the videos or system output to answer correctly (e.g., ``Is the sky dark during the night?'').
These attention check questions were used to determine whether participants were reviewing queries and making an effort to answer correctly.
If participants did not answer these attention checks correctly, we assumed they may not have been giving sufficient attention to the queries and task, and hence their data was not included for analysis.



\subsection{Procedure}

The experiment was run as a single-session online user study via Amazon Mechanical Turk (AMT).
The research was approved by the organization's institutional review board (IRB), and the participants were compensated based on a fixed rate per hour.
Average completion time depended on the assigned condition, with total time varying from approximately 25 to 55 minutes.

\begin{figure}[!t]
\centering
  \includegraphics[width=0.7\columnwidth]{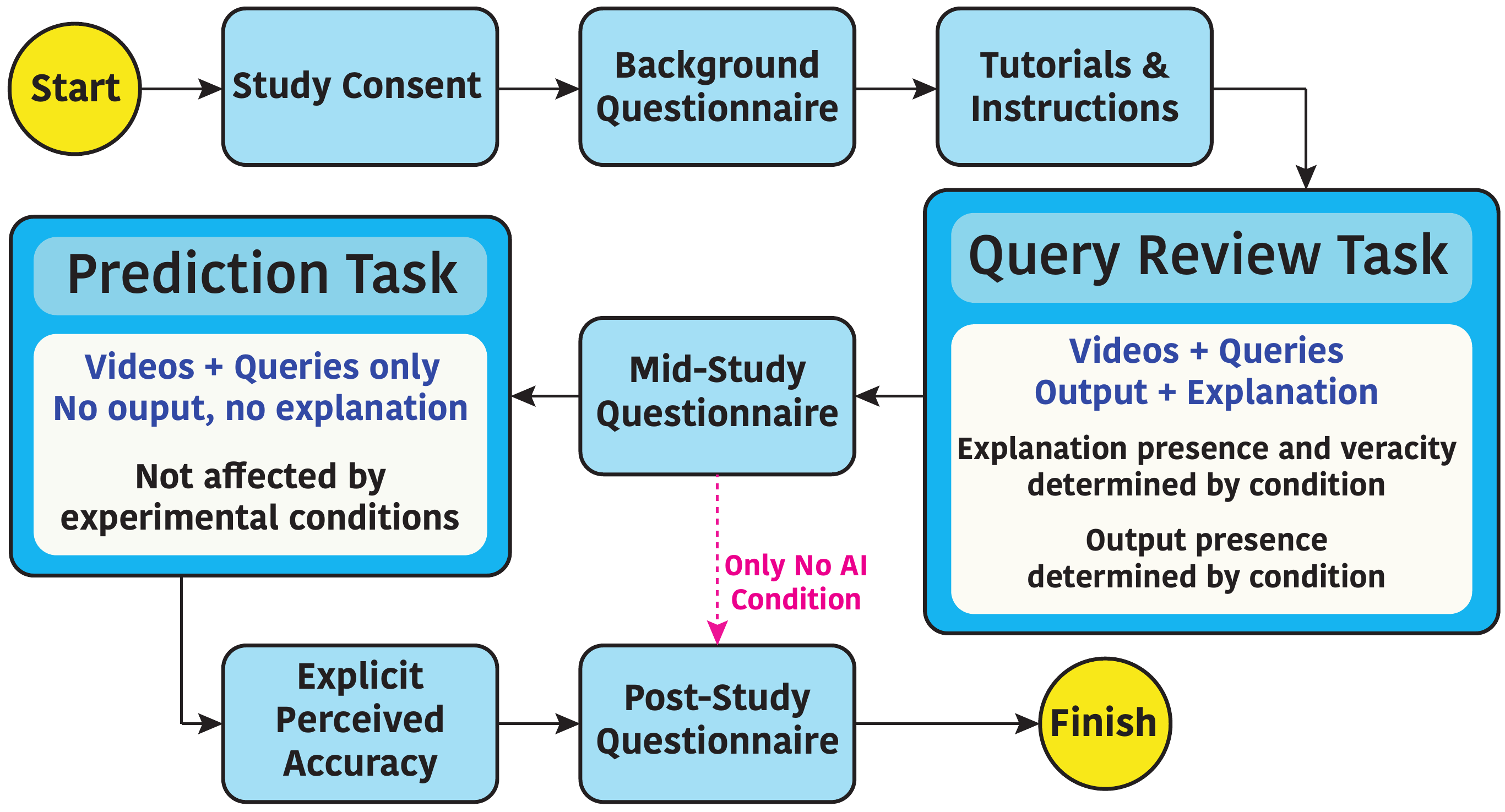}
  \caption{An overview of the study procedure. The \revTask{} is affected by conditions. All participants completed a version of the main task based on the assigned condition, and all did the same AI prediction task except the \noai{} condition.} ~\label{fig:procedure}
\end{figure}

Figure \ref{fig:procedure} provides a visual overview of the study procedure.
Participants first viewed a consent form, followed by completing a background questionnaire asking about participants' age, gender, education level, occupation, and knowledge and understanding of AI and ML.
After the questionnaire, participants were given a brief tutorial of the system and instructions for the \revTask{}.
Tutorials and instructions were crafted based on the assigned condition.
In order to make the tutorials more engaging, the task was described in the context of a cooking competition scenario and assessing a system that could help in contest judging.

The participants then completed the \revTask{}, which consisted of 20 queries.
In addition to being asked to determine the correct answer for each query, participants in the \fullXai{} and \randomXai{} conditions were also asked to rate whether they found the explanations helpful (see Figure~\ref{fig:interface}e).
After participants confirmed their response for each query answer, we showed the correct answer to provide feedback and to allow participants a clear indicator of whether the system's answer matched the true answer.
After showing the correct answer, the system allowed participants to continue to the next trial.
The users could spend as much time on each trial as they desired.

After the \revTask{}, participants took a short questionnaire about how much they used the different components and features of the interface and how helpful they found them.

Participants then continued to the \textit{prediction task}, which was not affected by the experimental condition.
A new set of instructions were given before the 20 prediction trials (again with no time limit).
No feedback about correct answers was available during the prediction task.
Note that participants in the \noai{} condition did not complete the prediction task since they did not observe any model output, so there was no model for them to predict.

After the prediction task, participants were asked to  explicitly estimate the accuracy as well as answer various additional questions asking for general feedback about the experience and system.

\subsection{Participants}

We recruited a total of 160 AMT workers, with 40  workers per condition.
After removing the data from the users who failed  quality checks (described in the following section), we reached a total of 38, 40, 40, and 38 workers for \fullXai{}, \randomXai, \noXai{}, and \noai{}.
We did not reject any HITs that were submitted for the study, so all  participants were paid regardless of outlier or quality removal.
There were no significant difference with participants' age distribution among conditions (M = 41.26 and SD = 11.37).
Also, 54 of the total participants were male and 64 female.
The gender distribution among the conditions were not significantly different.


\section{Results}
\label{sec:Results}
We analyzed the results from the study based on the measures from both performance and prediction tasks, as well as from questionnaires.

\begin{table}[!t]
    \resizebox{.8\linewidth}{!}{
    \setlength{\tabcolsep}{15pt}
    \renewcommand{\arraystretch}{1.5}
    \begin{tabular}{|c|c||r|}
         \hline
         \rule{0pt}{2ex}
         \multirow{2}{*}{\textbf{Time per Task}} & Main Effect &
         $\chi^2(3,150) = 34.7$ ($p<0.001$)  * \\ \cline{2-3} \rule{0pt}{2.5ex}   
         & \multirow{2}{*}{Post-hoc Test} & \noXai{} vs. \textbf{\fullShort{}} ($p<0.001$) * \\ 
         && \noai{} vs. \textbf{\fullShort{}} ($p<0.001$) * \\
         && \textbf{\randomShort{}} vs. \noXai{} ($p<0.001$) * \\
         \hline
         
         \rule{0pt}{2.5ex}   
         \multirow{1}{*}{\textbf{Average Error}} & Main Effect &
         $ \chi^2(3,150) = 28.8$ ($p<0.001$) * \\ \cline{2-3} \rule{0pt}{2.5ex}
         & \multirow{2}{*}{Post-hoc Test} & \randomShort{} vs. \textbf{\fullShort{}} ($p<0.001$) * \\
         && \noXai{} vs. \textbf{\fullShort{}} ($p<0.05$) * \\
         && \noai{} vs. \textbf{\fullShort{}} ($p<0.01$) * \\
         && \randomShort{} vs. \textbf{\noXai{}} ($p<0.001$) * \\
         && \randomShort{} vs. \textbf{\noai{}} ($p<0.05$) * \\
         \hline
         
    \end{tabular}}
    \vspace{0.5cm}
    \caption{The table \textbf{only} includes significant post-hoc results. For the post-hoc comparisons, the condition with the better performance is presented in bold. We used a Kruskal-Wallis non-parametric test for the main effect and a Wilcoxin post-hoc test for pairwise comparison.}
    \label{table:performanceResults}
\end{table}

\begin{figure}[!t]
    \centering
    \begin{subfigure}{0.49\linewidth}
        \centering{\includegraphics[width=0.7\linewidth]{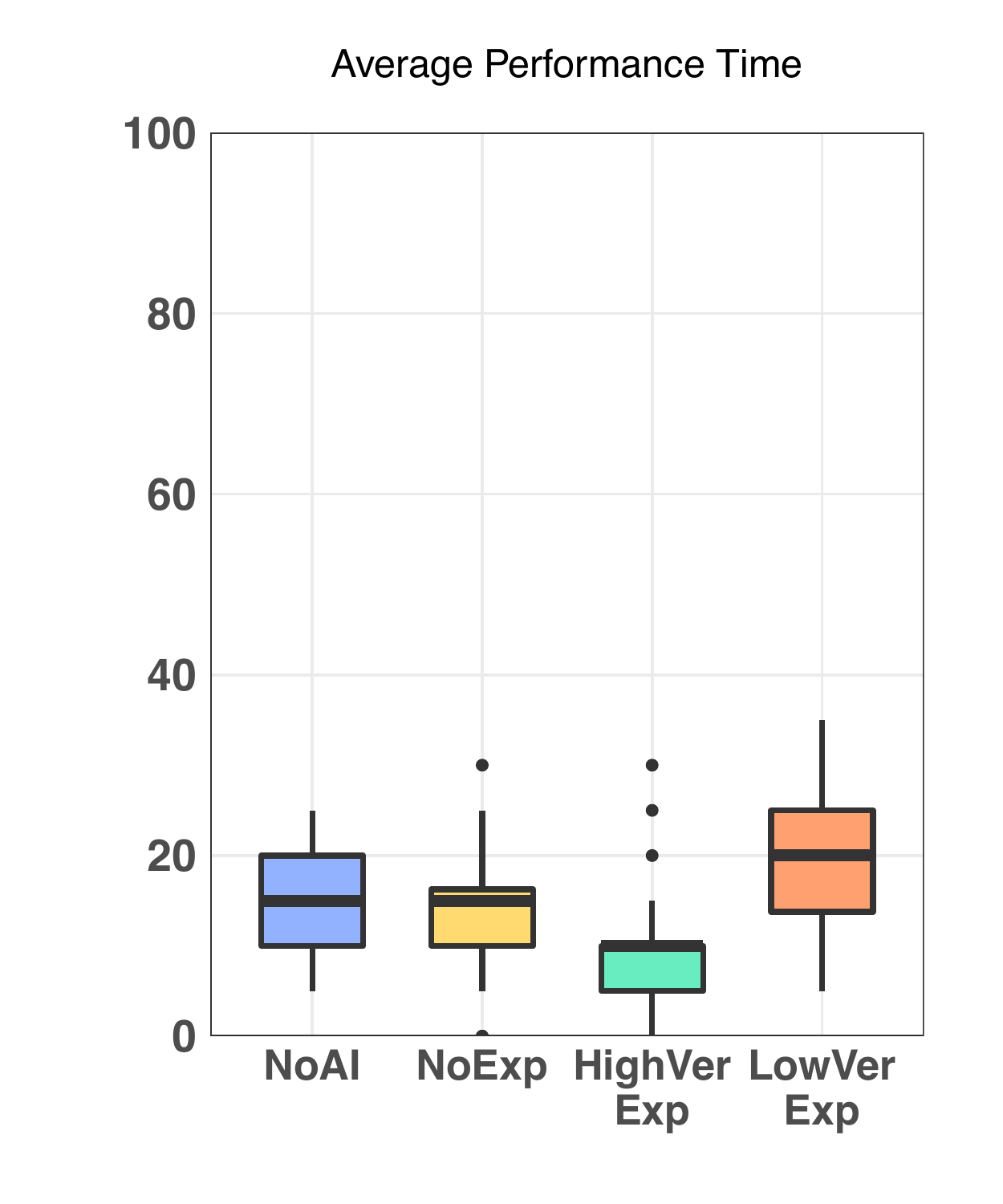}}
        \captionsetup{justification=centering}
        \caption{\textbf{Average Error per Task\\ (Percentage)}}
        \label{fig:perfCorrRQ0}
    \end{subfigure}
    \hspace{-2cm}
    \begin{subfigure}{0.49\linewidth}
        \centering{\includegraphics[width=0.7\linewidth]{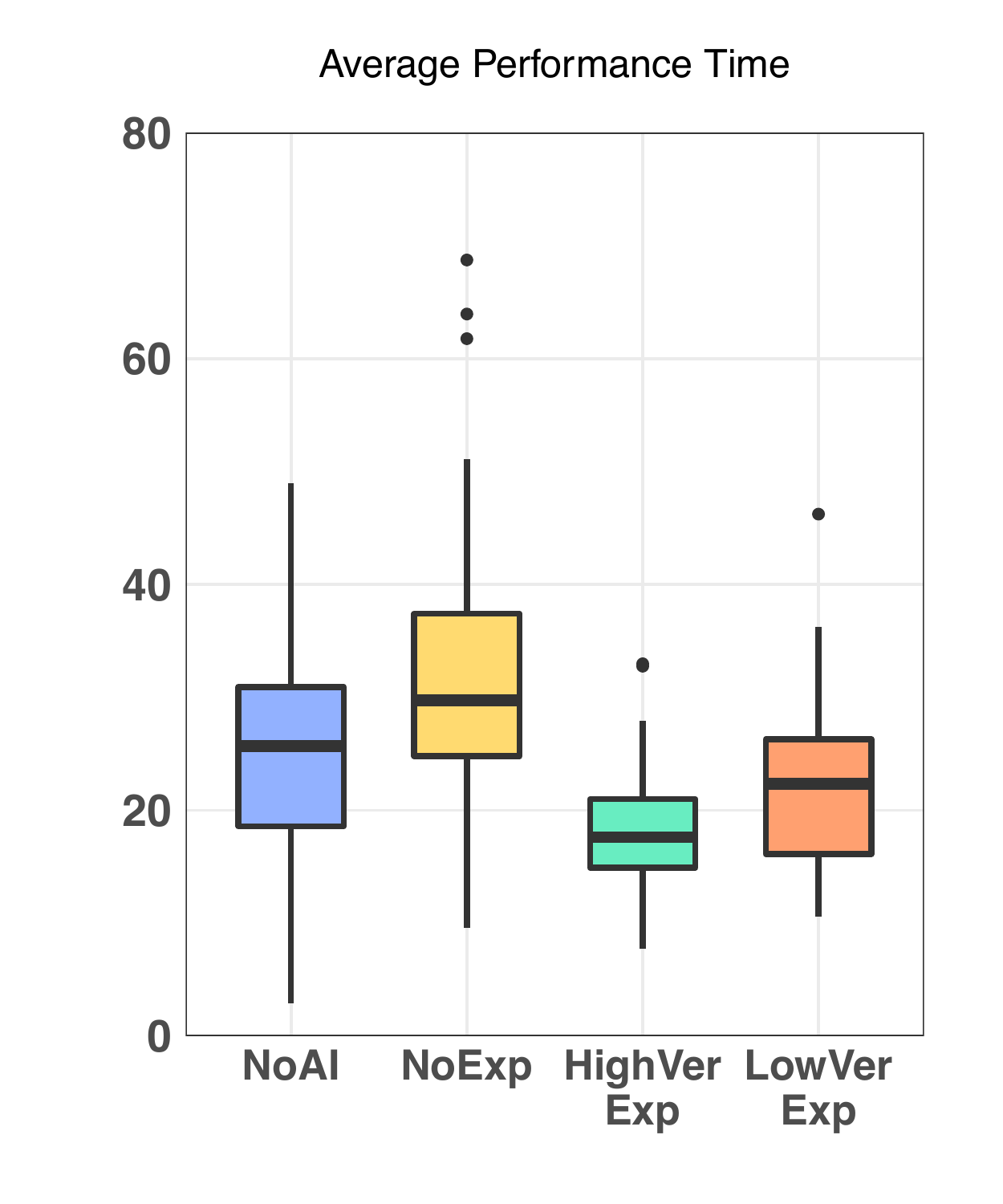}}
        \captionsetup{justification=centering}
        \caption{\textbf{Average Performance Time per Task (Seconds)}}
        \label{fig:perfTimeRQ0}
    \end{subfigure}
    \caption{User performance measures among the four conditions. Lower scores for both measures indicate better performance, i.e., lower errors and less performance time per each trial. 
    }
    \label{fig:performance-plots}
\end{figure}

\subsection{Pre-processing and Data Cleaning}

Before analysis, we conducted data filtering to account for outliers and potential quality issues common in crowdsourced user studies~\cite{paolacci2010running, hara2018data}.
Data was filtered based on task completion time and errors.
Completion was measured for each trial, and we first removed outlier trials based on time following the 1.5xIQR rule within each condition (i.e., we calculated $ \pm 1.5$xIQR  of every trial, and only included the results that fell within this range).
Secondly, we again applied 1.5xIQR outlier filtering for performance error based on the percentage of times each participant answered a query wrong.
From the outlier and quality filtering, two participants were discarded for having extremely poor performance (i.e., answering all tasks wrong, and having 85\% of trials as high time outliers).
Of the remaining participants, 1.25\% were removed as outliers.

\begin{table}[!t]
    \resizebox{0.8\linewidth}{!}{
    \setlength{\tabcolsep}{10pt}
    \renewcommand{\arraystretch}{1.5}
    \begin{tabular}{|c|c||r|}
         \hline
         \rule{0pt}{2ex}
         \multirow{2}{*}{\textbf{User Agreement}} & Main Effect &
         $F(2,150) = 22.67$ ($p<0.001$)  * \\ \cline{2-3} \rule{0pt}{2.5ex}   
         & \multirow{2}{*}{Post-hoc Test} & \randomShort{} vs. \textbf{\fullShort{}} ($p<0.001$) * \\ 
         && \noXai{} vs. \textbf{\fullShort{}} ($p<0.01$) * \\
         && \textbf{\noXai{}} vs. \randomShort{} ($p<0.05$) * \\
         \hline
         
         \rule{0pt}{2.5ex}   
         \multirow{1}{*}{\textbf{Exp Usefulness}} & Main Effect  (pairwise) &
         $F(1,76) = 4.18$ ($p<0.05$) * \\
         \hline
         
    \end{tabular}}
    \vspace{0.5cm}
    \caption{The results from user agreement with the system and explanation usefulness. 
    We used an independent one-way ANOVA to find the main effect and a Tukey HSD test for pairwise comparison.}
    \label{table:agreementResults}
\end{table}

\begin{figure}[!t]
    \centering
    \begin{subfigure}{0.49\linewidth}
        \centering{\includegraphics[width=0.65\linewidth]{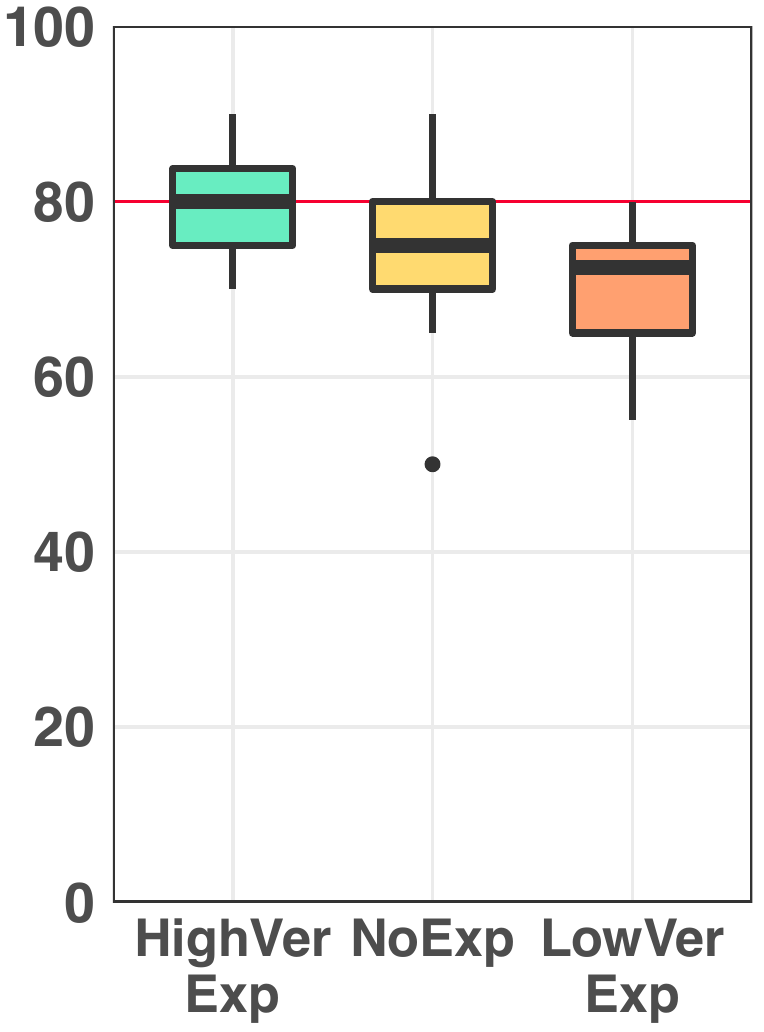}}
        \captionsetup{justification=centering}
        \caption{\textbf{User agreement \\with system's answer (Percentage)}}
        \label{fig:agreement}
    \end{subfigure}
    \hspace{-2cm}
    \begin{subfigure}{0.49\linewidth}
        \centering{\includegraphics[width=0.65\linewidth]{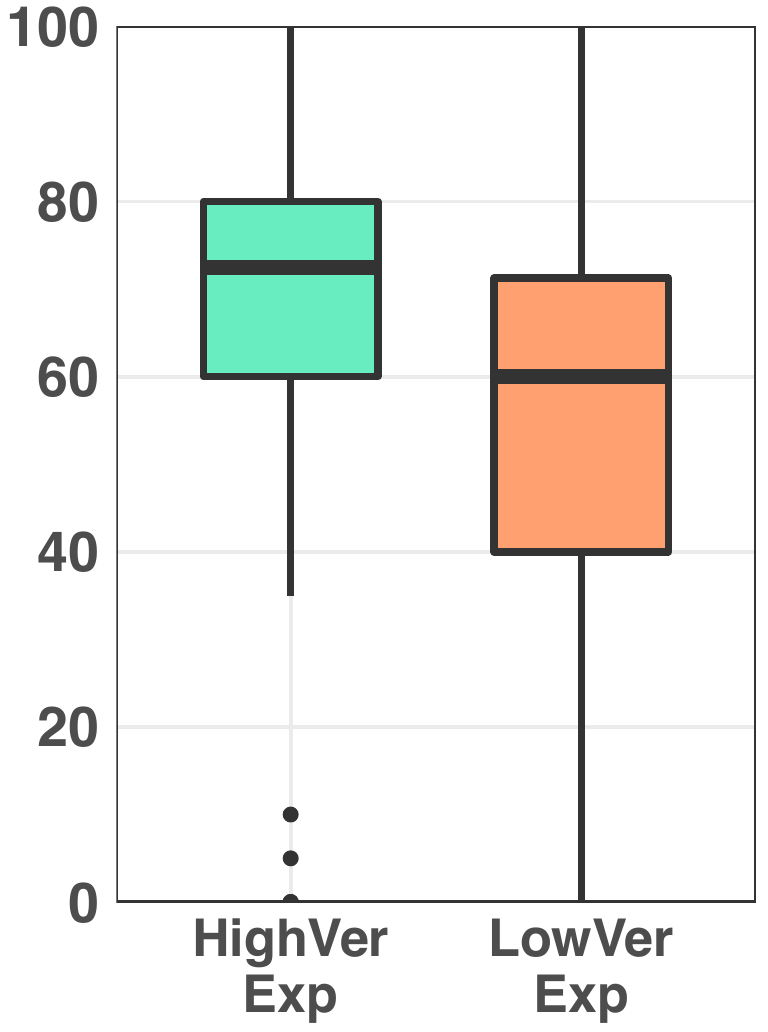}}
        \captionsetup{justification=centering}
        \caption{\textbf{Explanation usefulness \\(Percentage)}}
        \label{fig:usefulness}
    \end{subfigure}
    \caption{Additional measures from the \revTask{} for the explanation conditions. The left plot shows the percentage of times the participant's answer matched the system's answer. The horizontal red line represents the actual controlled accuracy of system answers. The right plot shows the number of times users found explanations helpful in the explanation conditions. Note that these measures were only possible for the conditions shown (i.e., participants cannot agree with the model in the \noai{} condition or rate explanation usefulness without explanations).}
    \label{fig:RQ1}
\end{figure}

\subsection{Query Review of Performance Results}

We have two main goals for performance assessment.
First, we want to evaluate whether or not the XAI system supports user performance improvements by providing AI answers both with and without explanations.
This goal involves comparisons of the \noai{}, \noXai{}, and \fullXai{} conditions, where \noai{} provides a reference for unassisted human performance.  Second, we want to understand how explanation veracity affects performance.  This goal is based on comparisons among the \fullXai{}, \randomXai{}, and \noXai{} conditions.
For this purpose, the \noXai{} condition serves as a baseline to understand the effects of different types of explanations while preserving consistency in the same AI answers available among these three conditions.

For statistical analysis, the data for the performance measures did not meet the assumptions for parametric testing, therefore we used Kruskal-Wallis non-parametric tests for analysis of time and error comparing the four conditions.
We conducted one test for time and one for error.
Following significant results from the omnibus testing, we used Wilcoxon rank sum post-hoc tests for pairwise comparisons.
Table \ref{table:performanceResults} shows a summary of test outputs and significance for user performance results, and Figure \ref{fig:performance-plots} shows these results graphically.

For both time and error metrics, participants who used the \fullXai{} system had significantly better performance compared to both \noXai{} and \noai{} conditions.
This means that the availability of AI answers alone did not result in significant performance improvements, but significant improvements were detected when the proper explanations were also added.
This finding informs RQ1 and provides significant evidence that the designed XAI system is effective at enhancing user performance.

The effect of explanation veracity was also significant.
The \fullXai{} group was significantly more accurate (fewer errors) than the \randomXai{} condition.
Task completion time was not significantly different between these two conditions.
Another important finding is that the \randomXai{} condition had significantly more errors than the from \noXai{} condition.
Interestingly, performance in the \randomXai{} group was also significantly faster than \noXai{}.

These results serve as strong empirical evidence that appropriate explanations can be beneficial for user task performance with the assistance of an intelligent system.
In this case, AI support was only beneficial when accurate explanations were also included (that is, AI output was not significantly helpful without explanations).
Moreover, having inaccurate explanations resulted in worse performance than having no explanations. 


\subsection{Agreement with the System and Explanation Usefulness}

By comparing user responses from the \revTask{} to the system's answer for each query, we evaluated \textit{agreement} between the participant's decision and the system's suggested answer.
Note that this measure of agreement is only possible for conditions were the system provides an AI answer, and we reiterate that all conditions provided the same accuracy of given AI answers.
Thus, we compared agreement percentages among \fullXai{}, \randomXai{}, and \noXai{} conditions that included the model output for the query.
We used a one-way independent ANOVA with Tukey HSD post-hoc testing.
The main effect was significant, and all pairwise comparisons among the three conditions were significant.
Table \ref{table:agreementResults} shows the test results, and Figure~\ref{fig:agreement} shows agreement results graphically.


Results show the highest level of agreement in the \fullXai{} condition while \randomXai{} was the worst.
Consider the \noXai{} condition as reference point, participants decided on answers aligning with the system's answers at a rate close to the 80\% simulated system accuracy.
Along with the performance results, this suggests that the appropriate explanations allowed participants to correctly identify system errors and disagree with the system when the system was wrong.
Participants in the \randomXai{} condition showed overall lower agreement with the system's output, which suggests lower trust in the AI output even though the fraction of correct AI answers was the same.
This suggests that poor explanations led participants to disagree with the system even in cases in which the AI answers were correct. 
This outcome is important as user disagreement with the system relates to perception of system accuracy and is an indication of reduced trust in the system.


To provide a clearer mapping of this agreement behavior to perception of the explanations themselves, we also evaluated user's responses from the secondary question from each trial of the \revTask{} (i.e., \textit{``Was the explanation Helpful?''}) for \fullXai{} and \randomXai{} conditions.
For analysis of overall perceived usefulness of explanations from the entire \revTask{}, we calculated the percentage of explanations rated as useful by counting trials where participants indicated responses of \textit{agree} or \textit{strongly agree} for the question about usefulness.


Results for this metric met the assumptions for parametric testing after applying a $x^2$ transformation.
A one-way independent ANOVA found a significant difference showing participants found low veracity explanations less useful.
Table \ref{table:agreementResults} shows the results.

\subsection{Mental Model and Perception of Accuracy}

As described in the Experimental Design section, a \textit{prediction task} was used to measure user's mental model and perception of accuracy in the XAI system based on the question (i.e., ``What would the system's answer to this query be?'' and ``Would the system's answer be correct?'').
Participants completed this task in the three conditions where participants had AI output (\noXai{}, \fullXai{}, and \randomXai{}).
We evaluated prediction accuracy (i.e., rate of correctly guessing the model's answer) with one-way independent ANOVA, which did not show a significant difference ($F(2, 115) = 0.64$).
This means we did not find evidence that explanation veracity and presence affected user ability to accurately predict the system's AI output.



In addition to analyzing correct predictions that matched the system's output, we also analyzed an implicit measure of perceived model accuracy based on the percentage of instances where participants predicted that the system would give correct answers. 
The implicit user perceived accuracy did not show a significant difference among the conditions ($F(2, 115) = 0.40$).
We also explicitly asked participants to report a numerical estimation of the accuracy as a percentage.
An ANOVA also did not detect a significant effect across the conditions ($F(2,115) = 1.9$).

Thus, while veracity of explanations did help user assessment of system results enough to help significantly improve user performance, the understanding of the model was not affected enough to provide a significant effect on prediction of model outputs or accuracy in this task.

\section{Discussion}
\label{sec:Discussion}
In this section, we discuss the results, implications for XAI design, and limitations of the research.

\subsection{Results Interpretation}

In this study, our main goal was to understand how the explanation veracity (i.e., the accuracy and capability of the explanations to justify the model's output) can affect user performance.
We also wanted to understand how explanation veracity and presence would affect user's agreement with the system output, perception of the accuracy, and mental model of an XAI system.
In order to make this comparison and generalize the findings for other XAI systems, we utilized a customized XAI system and analyzed if the explanations generated by the model can improve user performance.
The results found evidence that users of this system had significantly better performance (i.e., were faster with less error).
To test the effects of veracity, we compared three conditions of \textit{high} and \textit{low} veracity explanations and \textit{no} explanations.
For this comparison, while the accuracy of the system answers remained constant (80\%) across all conditions, the accuracy of explanations was controlled to be high (as determined by the model) for \fullXai{} and low (crafted specifically to be inaccurate) for \randomXai{} conditions.
Our results show user performance was significantly better with the model-generated high veracity explanations compared to low veracity explanations.

From another perspective, performance times were always significantly faster with \textit{any} explanations than without, though errors were significantly lower with high-veracity explanations while the addition of low-veracity explanations resulted in faster performance and more errors.
Users were more error prone with low-veracity explanations even compared to cases with no explanations at all. 
This phenomenon suggests that the presence of explanations can encourage users to mistrust the system even when it is right.
Another interpretation is that users found low veracity explanations less meaningful, resulting in their loss of trust and confidence in system outputs.
This may have demotivated participants or promoted rushed and reduced effort.
Additional studies are required in order to better understand the implications of effective and poor explanations.
In either case, the results clearly indicate the importance of being conscientious and careful with including explanations before understanding whether the explanations seem meaningful to end users.

From the agreement results, we can infer that the participants from \randomXai{} thought the system was wrong more times than the system actually was (see Figure~\ref{fig:agreement}).
This happened while the users in this condition also made more errors in their task compared to the other two conditions.
Given that the output accuracy of the system was constantly 80\% across the conditions with AI answers available, the difference in agreement and error level is caused by explanation veracity.
This means that given two systems with the same model but different explanations, users with less accurate (or even less ``human meaningful'') explanations might misunderstand the system, which is expected to cause mistrust in its output.
Furthermore, users of such a system might falsely disagree with the system for the sole reason that the explanations do not make sense to them.

Another issues for consideration is whether users with \randomXai{} suffer from automation bias~\cite{cummings2004automation,zaroukian2017automation,sauer2016experience,sundar2019machine,bussone2015role}.
Automation bias is often times referred to as over-reliance on automated outputs~\cite{bussone2015role}, over-trusting output from automated systems even when it is wrong~\cite{layton1994design,sundar2019machine}.
The results show participants with \randomXai{} were faster and made more errors compared to \noXai{} based on the results of the \revTask{}.
At first glance, this result could look like automation bias since users were faster in agreeing with the system even though the system justifications were wrong, but the user agreement results refute such a conclusion.

In this experiment, the performance measure captures when users both agree with the model when it is correct and disagree when it is wrong---in other words, correctly identifying when to agree or disagree with the system (Bussone et al.~\shortcite{bussone2015role} refer to this as \textit{right} decisions).
In contrast, the user agreement measure captures when users agree with the model's answer regardless of whether it was correct or incorrect.
Since users with \randomXai{} had worse performance error and lower agreement, this means they were not over-relying on the system, but rather they were under-relying on its outputs.

In other words, given that participants in the \randomXai{} condition were significantly faster than the \noXai{} condition, they may have quickly disagreed with the system answer out of mistrust of the system's answers, as indicated by both lower agreement and greater error with low-veracity explanations.
With poor explanations, participants lost trust in the system, and they were so untrusting that they were \textit{quick} to discount the system's answer and instead were more likely to answer \textit{counter} of the system's answer.
This, as it turns out, is the \textit{opposite} behavior as we would expect from automation bias.
Since the model was usually correct (80\%) and participants were more likely to disagree due to mistrust, the \randomXai{} group ended up with worse performance overall.



\subsection{Implications for XAI Design and Evaluation}

The findings from this experiment are important for designers who are interested in providing explanations for end users of intelligent systems. 
In our previous work~\cite{nourani2019effects}, our main finding indicated that explanations that are not meaningful to the users lead to user underestimation of system accuracy.
Previous work~\cite{yin2019understanding} showed that underestimation of system accuracy can directly influence user trust and can cause mistrust.
Hence, combining these two findings, we can conclude that non-meaningful explanations can significantly affect user trust.
Although the results of our current study did not show significant differences for user perception of accuracy across the conditions, the participants who observed low veracity explanations did not rely on the system, found explanations less meaningful, and more importantly, they even tended to disagree with the system even when it was correct on the prediction.
These studies highlight the importance of both explanation meaningfulness and veracity together for affecting how much users rely on its predictions.
Both can affect human-task performance and trust in the automated system.
Especially since many XAI projects do not include human evaluations~\cite{adadi2018peeking,mohseni2019survey}, a major concern is that system designers might include explanations without first evaluating them, as developers might expect the presence of any explanations to help improve user performance.
However, the results of this study suggest that veracity and perceived usefulness of explanations also play an important role in how added explanations affect user performance and behavior.
Human evaluation is, thus, essential to make sure they are understood and working as intended.
While such a claim is perhaps not surprising for an HCI audience, such empirical results are important for XAI system designers.

The findings also highlight the importantance for consideration of the level of task complexity and user domain knowledge.
For example, in explainable decision-support systems, it is critical that users, who may be domain experts such as medical professionals, security experts, or emergency responders, do not make serious or fatal errors in decision making.
Disagreement with such systems could cause serious problems or even be fatal.
For instance, given such explainable system with a high accuracy of 96\%, providing low veracity explanations could cause a less experienced doctor to lose trust in the system's output and disagree with the output, which in this case is a diagnosis.
These situations where the user disagreed with the diagnosis falsely could lead to a false diagnosis.
Similarly, a more experienced user could lose trust and stop using an explainable system completely after observing the low veracity explanations.
In this case, the user's task-performance might significantly decrease to a level compared to when there is no system available (without automation).
In an example from prior literature, Bussone et al.~\shortcite{bussone2015role} designed a study to test trust and reliance on explanations in a clinical decision support system, which was targeting less experienced healthcare practitioners without a specific specialty knowledge.
Their results showed that more detailed explanations would cause over-reliance while little information can cause self-reliance on the automated system.
Explanation meaningfulness and veracity might also affect a user's decision-making accuracy over time.
This situation and similar situations could be avoided if the system designers approach with cautious when designing and providing explanations in an intelligent system.

Another important consideration is designing systems that would provide the option of improving system predictions and outcomes through user feedback.
In these situations, both over-reliance and under-reliance can cause problems with changes of the model over time and its accuracy.
With mistrust and under-reliance, users might assume the system is wrong even with the cases when it is not, thus causing them to provide feedback that would falsely change the model.
Some of this feedback, depending on how much and when they are implemented in the system, could cause the model prediction and performance to change drastically.
It could later affect user's decision accuracy, especially if they are new users and trying to establish trust, or if the users already over-rely on the model's outcomes.
As our results show, explanation veracity can affect user reliance, and hence, designers of such systems should consider that low explanation veracity might eventually affect the accuracy and quality of their model.

\subsection{Limitations and Opportunities}

Although the study found significant effects of explanation types on human-machine task performance, the study did not detect evidence that the addition of explanations helped participants to develop a mental model accurate enough to improve responses in the prediction task.
Taken together, these results indicate that although the XAI system was able to be understood sufficiently to judge whether the output was accurately meaningful and to determine when to rely on the output, it was not sufficient for forming an overall understanding of why the system was right when it was right or why it was wrong when it was wrong. 
We believe this is likely related to limitations on study duration and number of trials in the query performance phase.
We aimed to keep the study time low (around 30 minutes) to accommodate online testing, but this also limited the number of trials a participant had to interact with a system. 
With the chosen design, we suspect participants did not have sufficient opportunities to observe incorrect model results.
We suspect longer periods of interaction would have been needed to develop any type of meaningful mental model of the XAI model. 
It may be necessary to consider extended study times and observe more model weaknesses.

For future consideration, it may also be interesting to consider how user expertise and experience could affect their performance and how they interpret explanations in intelligent systems.
The task in our study was intentionally designed such that completion did not require any specific domain knowledge.
In some cases, users with domain knowledge might have different performance and mental model of the underlying AI system.
Experience and understanding of ML/AI can also affect how users use a system and how they build a mental model of how it is working.
Using AMT for participant recruiting, our experiment did not target participants with expertise or significant knowledge of AI.
It would be interesting to conduct similar studies that also accounts for how the level of expertise in ML/AI affects user performance.

As advances in research contribute new knowledge of potential benefits and concerns of explainable systems, it is important to consider implications across different types of intelligent systems for different purposes.
Testing the concepts we explored in this paper with different systems or with environments that require proficiency in a specific domain can bring about a deeper understanding and a more generalizable knowledge for the future of XAI systems.
Our XAI system in this study was operating with component-level explanations at a high level -- geared for end users -- rather than low-level explanations with model details.
While these results are generalizable for component-level explanations, future studies need to test these findings on low-level explanations.
In other words, we expect generalization to depend on the understandability of the task and explanations, and more studies would be needed to assess different depth of explanation and alternate data contexts or explanation types.
With continued developments in AI and machine learning approaches as well as with the evolution of public understanding and perception of intelligent systems, we argue for the importance of continuing empirical studies over a broad set of applications, models, and contexts.

In addition, as part of building a strong empirical basis for our knowledge of the effects of explanations on human understanding and behavior, it is essential to study a variety of different forms of user tasks.
We are interested in exploring more complex user tasks that involve more freedom and decision-making in how to use the AI and explanations to complete the tasks.
While more open-ended user tasks increase variance in user activity and can introduce challenges for controlled comparisons of conditions, they can also serve as more realistic scenarios for study and can allow us to study when users are willing to rely on the system over time and when frustrations might arise.


\section{Conclusion}
\label{sec:Conclusion}
Explainable AI systems aim to reduce problems of many of the black-box models for end users to develop appropriate understanding and trust in algorithms.
However, if explanations lack accuracy or are not able to meaningfully describe the model, they might result in negative or unintended user behaviors.
Motivated to understand the effects of such explanations on user behaviour, the presented user study evaluated differences in user performance, agreement, perception of accuracy, and mental model.
The findings demonstrate that the quality of explanations can have significant---and potentially opposite--effects on effectiveness and utility based on human interpretability.
This research provides a clear case for the importance of evaluating explanation design through human-subjects evaluation in addition to emphasizing model accuracy and computational measures of explanation soundness alone.
An explainable model whose explanations are not accurately describing its logic can cause several problems for its end-users, such as mistrust, lower performance, and lack of proper understanding of the model.


\begin{acks}
This work was supported by the DARPA Explainable Artificial Intelligence (XAI) Program under contract number
N66001-17-2-4032.
\end{acks}

\bibliographystyle{ACM-Reference-Format}
\bibliography{Bibliography}

\end{document}